\documentclass{aastex631}

\usepackage{siunitx}

\begin{document}

\title{Ion kinetics of plasma interchange reconnection in the lower solar corona}

\author[0000-0002-6809-6219]{Vladimir Krasnoselskikh}
\affiliation{LPC2E, CNRS-University of Orl\'eans-CNES, Orl\'eans, France}
\affiliation{Space Sciences Laboratory, University of California, Berkeley, USA}

\author[0000-0001-8543-9431]{Arnaud Zaslavsky}
\affiliation{LESIA, Observatoire de Paris, Meudon, France}

\author[0000-0001-8823-4474]{Anton Artemyev}
\affiliation{Departement of Atmospheric Studies, University of California, Los Angeles, USA}

\author[0000-0001-5315-2890]{Clara Froment}
\affiliation{LPC2E, CNRS-University of Orl\'eans-CNES, Orl\'eans, France}

\author[0000-0002-4401-0943]{Thierry Dudok de Wit}
\affiliation{LPC2E, CNRS-University of Orl\'eans-CNES, Orl\'eans, France}
\affiliation{ISSI, Bern, Switzerland}

\author[0000-0003-2409-3742]{Nour E. Raouafi}
\affiliation{Johns Hopkins Applied Physics Laboratory, Laurel, USA}

\author[0000-0001-6427-1596]{Oleksiy V. Agapitov}
\affiliation{Space Sciences Laboratory, University of California, Berkeley, USA}

\author[0000-0002-1989-3596]{Stuart D. Bale}
\affiliation{Space Sciences Laboratory, University of California, Berkeley, USA}
\affiliation{Physics Department, University of California, Berkeley, USA}
\author[0000-0003-1138-652X]{Jaye L. Verniero}
\affiliation{NASA Goddard Space Flight Center: Greenbelt, MD, US}
 


\begin{abstract}
The exploration of the inner heliosphere by Parker Solar Probe has revealed a highly structured solar wind with ubiquitous deflections from the Parker spiral, known as switchbacks. Interchange reconnection (IR) may play an important role in generating these switchbacks by forming unstable particle distributions that generate wave activity that in turn may evolve to such structures. IR occurs in very low beta plasmas and in the presence of strong guiding fields. Although IR is unlikely to release enough energy to provide an important contribution to the heating and acceleration of the solar wind, it affects the way the solar wind is connected to its sources, connecting open field lines to regions of closed fields. This ”switching on" provides a mechanism by which plasma near coronal hole boundaries can mix with that trapped inside the closed loops. This mixing can lead to a new energy balance. It may significantly change the characteristics of the solar wind because this plasma is already pre-heated, and can potentially have quite different density and particle distributions. It not only replenishes the solar wind, but also affects the electric field, which in turn affects the energy balance. This interpenetration is manifested by the formation of a bimodal ion distribution, with a core and a beam-like population. Such distributions are indeed frequently observed by the Parker Solar Probe. Here we provide a first step towards assessing the role of such processes in accelerating and heating the solar wind.

\end{abstract}



\section{Introduction} \label{sec:intro}

The exact mechanisms by which the slow and fast solar winds are generated are still actively debated. The distinction between the two wind regimes is primarily related to important differences in the characteristics of the magnetic field in the source regions, in particular the structure and dynamics of the coronal magnetic field. Early extreme-ultraviolet (EUV) and X-ray observations have shown that the coronal magnetic field consists of closed loops and open coronal hole topologies \citep{Zirker1977}. Coronal holes (CHs) are low density areas that appear darker in X-ray and EUV images. They are prominent at the solar poles during solar minimum, but occur also at lower latitudes \citep{Chiuderi-Drago1999}. The magnetic field flux within CH regions is not balanced, as one polarity often dominates, leaving the magnetic field lines open to interplanetary space in the upper coronal altitudes. The large-scale, fast, cool, and homogeneous solar wind originates from coronal holes \citep{Geiss1995, McComas2002}, propagating from the low-beta corona to the heliosphere along open field lines. CHs are well known to generate fast solar winds in polar regions. At lower latitudes, however, they are the source of moderately fast winds. The situation is different for the slow solar wind, which in the corona is significantly hotter, and exhibits considerably higher variability. Slow winds are generally co-located with the global streamer belt \citep{Gosling1997, Zurbuchen2002}. Numerous observations of transient X-ray and EUV brightenings at the boundaries of the CHs \citep{Kahler1990, Kahler2002, Kahler2010, Bromage2000, Madjarska2009, Subramanian2010} suggest that the slow wind generation is closely related with the dynamics of CH boundaries. The authors of \citet{Wang1990} proposed that the slow wind emerges from the CH boundaries due to super-extension of the magnetic field lines. Recent observations by Parker Solar Probe \citep[PSP;][]{Fox2016,PSP_Raouafi2023} provide a strong additional argument in favor of the emergence of the slow wind from these regions \citep[i.e., CH boundaries;][]{Fargette2021, Bale2021}.

Many basic properties of CHs have been studied using observations from the Solar Ultraviolet Measurements of Emitted Radiation \citep[SUMER;][]{Feldman1998} spectrometer on board the Solar and Heliospheric Observatory \citep[SoHO;][]{Domingo1995}. Their relation to open magnetic fields is established \citep{Wiegelmann2004, Meunier2005,  Zhang2006, Zhang2007, Yang2009a, Yang2009b}. It was also shown that some lines undergo blueshifts, indicating plasma outflows within CHs \citep{Hassler1999, Peter1999, Xia2004, Tu2005Sci, Aiouaz2005, McIntosh2011}. Observations by HINODE \citep{Kosugi2007} have shown that the magnetic fields at CHs are non-potential. That is, in some regions with strong magnetic fields within CHs, current densities can be as large as in flaring active regions \citep{Yang2011}. Mid-latitude CHs can be ``isolated'' or connected to polar CHs. In the latter case, they are called equatorial extensions of polar CHs \citep[(EECH), e.g.][]{Insley1995}. Observations reveal that they rotate quasi-rigidly although the photosphere rotates differentially \citep{Timothy1975, Insley1995, Wang1994}. 
 
CH boundaries (CHBs) separate two types of regions with different configurations: CHs with open magnetic fields and the surrounding quiet Sun with closed magnetic fields. The presence of the shear flows around coronal holes and the magnetic field configuration where open and closed field lines are rather close create favorable conditions for the development of forced and spontaneous magnetic reconnection. This implies that magnetic reconnection should necessarily be present at CHBs, otherwise the rigid rotation of EECHs could not co-exist with differential rotating surrounding photospheric fields \citep{Wang1994, Fisk1999}.

Conventionally, the notion of magnetic reconnection suggests strong energy release during the reconfiguration of magnetic field lines, where the energy source is provided by the oppositely directed magnetic field components. It is supposed that these processes play an important role in energy balance. Observational manifestations of intense energy releases appear as bright points in soft X-ray and EUV images. Surprisingly, \citet{Kahler2002}, who studied the equatorial extensions of polar CHs observed with the Yohkoh Soft X-ray telescope, found no significant effect of bright points on the CHB evolution. On the other hand, \citet{Kahler2010}, using images from Skylab X-ray, noticed that X-ray bright points at CHBs play an important role in the expansion and contraction of CHs. 

Using EUV images from the Transition Region and Coronal Explorer (TRACE) and SoHO, \citet{Madjarska2009} investigated the evolution of CHBs at small scales. They found that small-scale magnetic loops (which appeared as bright points) play an important role in the evolution of CHBs but found no signature for a major contribution from these bright points to energy release during the evolution of the loops. \citet{Kahler2010} also focused on the changes in CHBs but found no signature of energetic jets in EUV images.

The standard picture of the structure and evolution of the low-beta corona is described by the widely accepted so-called `quasi-steady' model \citep[e.g.,][]{Antiochos2007}. According to this model, the evolution of the corona is represented as a slowly evolving series of stationary potential field expansions or MHD equilibrium states. Such a macroscopic description may be justified as an acceptable first-order approximation since the large difference between the Afv\'en velocity and the characteristic velocities of the photospheric motions: $|V_{A}| \sim 10^3 \times  V_{flow}$, that means that the photosphere may easily adjust to any externally applied stresses \citet{Heyvaerts1984}. It is important to note that the steady-state model does not exclude the presence of the transient reconnection processes in the vicinity of the polarity inversion regions. These models do not explicitly include any dynamics, they simplify the flows to streaming along open field lines, and cannot explicitly place closed-loop plasma on open field lines. Thus, the slow wind within the quasi-steady framework can be accounted for by averaging intrinsically intermittent sources. The solution proposed to explain the formation of the slow wind is based on assumption that it originates from open flux tubes at coronal hole boundaries that expand much faster than the inverse square of the radial distance \citep{Wang1994, Arge2000}. As a consequence of this overexpansion, the flow speed is slowed proportionally. While overexpansion effects can explain the slower wind speed and to some extent the differences in the frozen-in ionization states \citep{Cranmer2007}, it cannot account for the composition and elemental abundance profiles that are known to be very similar to closed-field plasma sources. 

A fundamentally different, statistically based model for the coronal magnetic field structure based on so-called interchange reconnection (IR) has been proposed in \citet{Gosling1997, Fisk1999, Fisk2001, Fisk2005, Fisk2006}. 

\begin{figure}[h!]
\centerline{\includegraphics[width=11cm]{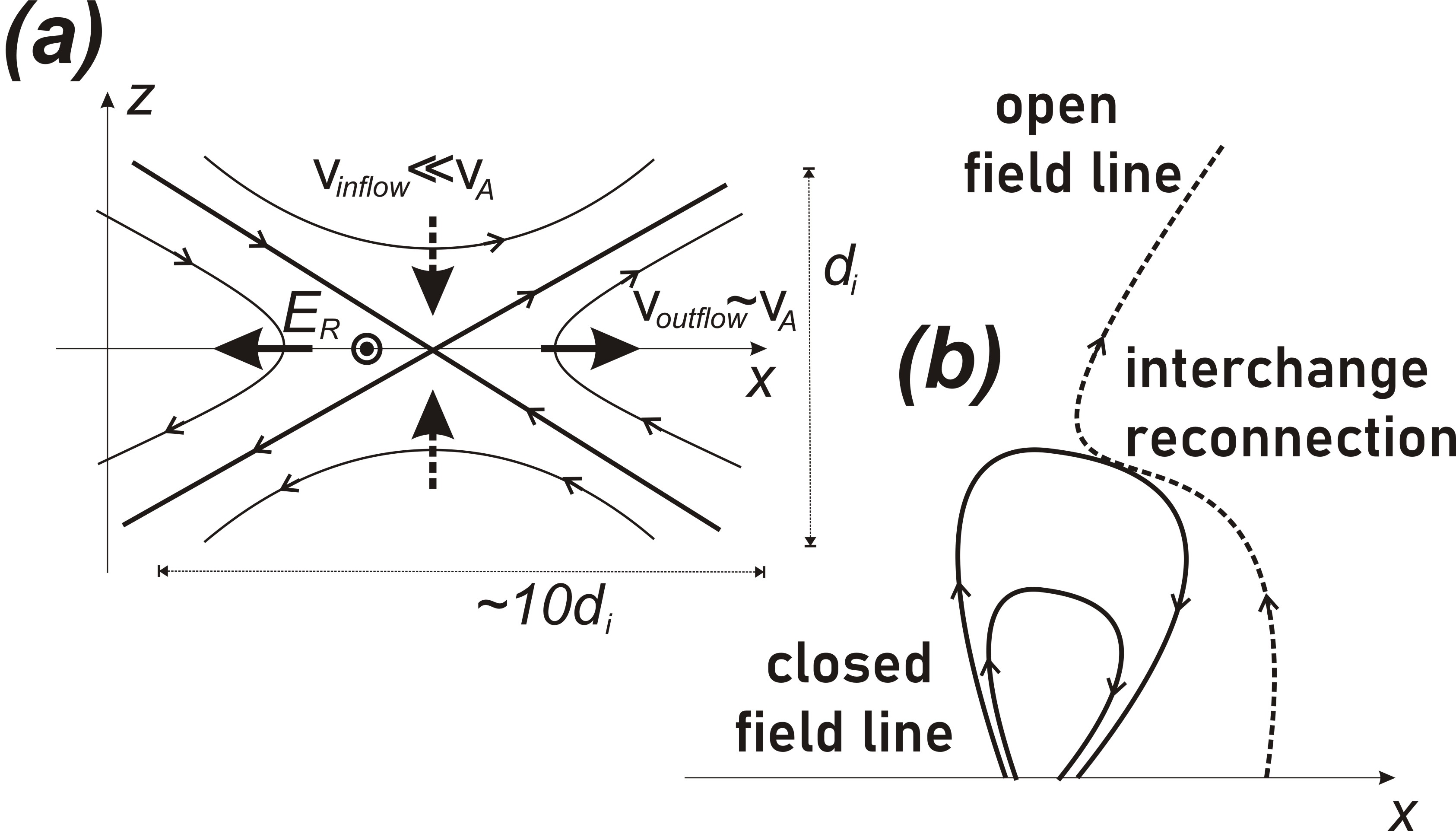}}
\caption{Schematic field of 2D geometry of magnetic field line reconnection (a) and interchange reconnection in the solar corona (b).}
\label{fig:scheme}
\end{figure}

The concept of IR process suggests that the magnetic reconnection involves the field line of some magnetic loop and the line open to the solar wind \citep[see][and a schematic in Figure~\ref{fig:scheme}(b)]{Crooker02,Fisk2001, Fisk2005, Priest&Forbes02,Aschwanden02}. Interchange reconnection is expected to deeply alter the magnetic field line topology \citep{Titov17,Pontin&Priest22} and lead to formation of the current sheet between open and closed magnetic field lines. Such plasma boundary-like structures formed within the reconnection region will then expand into the solar wind. This merging between closed and open fields thus  results in the release of plasma into the solar wind. As the height of the typical magnetic loop is supposed to be significantly smaller than the solar radius \citep{Aschwanden96,Aschwanden96:scaling}, this corresponds to the region of the dense low-$\beta$ (the thermal plasma pressure is significantly smaller than the magnetic field one) semi-collisionless plasma. 

The series of IR processes leads to diffusive transport of the open-field component throughout the corona, which can travel far into the region occupied by randomly oriented closed loops.  This diffusion model can be strongly supported by systematic convective motions, such as differential rotation, meridional flows, and granular convections as external drivers of the diffusion process. In such an approach, the random IR events determine a diffusion coefficient. Such diffusive transport of the open-field component may lead to a magnetic topology consisting of a highly complex mixture of open and closed fields generating disconnected coronal holes immersed deep within closed loop regions.

Photospheric motions create stresses that lead to the formation of the macroscopic current sheets around separatrix surfaces between the regions of open and closed field lines. The most important photospheric driving flows for the current sheet formation are those that inject helicity, see for example Figure 4 in \citet{Edmondson12}. These are the motions with streamlines that are parallel to constant contours of the radial magnetic field $B_{r}$ everywhere, which corresponds to pure helicity injection that is incompatible with quasi-steady evolution \citep{Edmondson10}. Relaxation of these stresses occurs through the formation of the macroscopic current sheets and IR. The characteristics of the process are determined by the local Lundquist number, which, in its turn, is determined by the resistivity.

 It is important to point out here that in a low-$\beta$ plasma, the input of the thermal plasma pressure to the pressure balance (even in the region of the magnetic field line merging) \citep[in the reconnecting current sheet, ][]{Priest85} is almost negligible. The current sheet (dynamical) equilibrium can then only be reached due to magnetic field shear. Thus, this current sheet should be force-free with ${\bf B}\approx const$, with the dominant role of field-aligned currents, whereas IR occurs in the presence of the strong guiding field and shear. Another important feature of such reconnection is its three-dimensional character: the reconnection-driven magnetic field reconfiguration results in the displacement of the footpoint of the open field line within the region of closed field lines \citep{Edmondson10,Edmondson12}. The plasma from closed magnetic field lines then penetrates into the open field lines and forms plasma-plasma boundaries separating hot dense and cold rarefied plasmas. Indeed, MHD simulations show that IR may form discontinuity-type  boundaries in the solar wind flows \citep{Burkholder&Otto19, Burkholder19}. 

 Kinetic processes responsible for the magnetic field energy dissipation with the violation of the frozen-in flux condition scale with ion inertial length, $d_i$ \citep{Cassak05,Yamada07}, that is orders of magnitude smaller than the characteristic length scales, $L$, of the typical structures such as magnetic arcs, i.e. in kinetic (collisionless or semi-collisional regime) processes of magnetic field energy dissipation occupy a quite small fraction of the space \citep{Yamada10:reconnection}. Taking this into account, one can speculate about the possible role of IR in magnetic field energy dissipation. The key characteristic of the reconnection is the reconnection rate, which is the rate of magnetic flux dissipation, or, equivalently, that of the reconnection electric field $E_R$ (such that $jE_R$ is the energy dissipation rate for the current density $j$) in the reconnection region (see schematic in Figure~\ref{fig:scheme}(a)). We take into account a supposed universality of the reconnection rate \citep{Liu22:reconnection_rate} and its weak dependence on the plasma $beta$ \citep{Birn10,Wilson16:cs,Drake21}. The dimensionless reconnection rate $c E_R /v_AB$ has been shown to be around $0.1$ for a quite broad class of plasma systems \citep{Liu17:reconnection_rate} (here $B$ and $V_A$ are background magnetic field and Alfv\'en speed). Therefore, we may use this rate to estimate an effective (collisionless) Lundquist number $S\sim 10^2\cdot L/d_i\gg 1$. As we expect the scale $L$ of the system (the IR region) to be much larger than ion inertial length, the collisional or semi-collisional \citep{Drake&Lee77} reconnection should have $S\gg 10^2$ (i.e., it should be much larger than the numerical Lundquist number for MHD simulations). Such reconnection should be most efficient for change of magnetic field topology \citep[similar to the interplanetary magnetic field reconnection with the Earth's dipole field, see][]{Merkin&Crooker08} without a significant contribution to the magnetic energy dissipation. This becomes important when the reconnection side moves to low-density altitudes with a strong decrease in $L/d_i$. Such conditions may appear in interplanetary space rather than in the corona.

One can therefore conclude that the energy release in such processes is negligible in terms of plasma heating or macroscopic acceleration. However, it opens the door to another important source of energy and particles for the solar wind, which will be the focus of our study, namely, the inter-penetration of plasmas: plasmas trapped in regions with closed field lines can penetrate into regions with open field lines and become a source of the solar wind.

As a result, the interchange model predicts that the slow wind originates from globally closed field regions with closed-loop plasma properties. The diffusive nature of the interchange model may naturally explain both the composition and the thick angular extent of the slow wind. As it was pointed out by \citet{Edmondson12}, inserting the open coronal hole flux tube deep inside the closed field regions requires a set of current discontinuities throughout the entire coronal volume, which contradicts the smooth quasi-steady magnetic field phenomenology. All these arguments lead to the conclusion that the reconnection dynamics in the vicinity of the coronal hole boundaries are the key to the generation of the slow solar wind. Such a change in magnetic field topology should result in the mixing of hot dense plasma of closed magnetic field lines and cold rarefied solar wind plasma, and the formation of plasma-plasma boundary, that may propagate to the solar wind and may be observed as a discontinuity \citep{Burkholder&Otto19, Burkholder19}.  \citet{Raouafi2023MRSW} found evidence for ubiquitous magnetic reconnection resulting in the production of small-scale jets (i.e., jetlets) throughout the base of the solar corona, in open and closed field regions.

\section{Description of the interpenetration of two plasmas}

The reconnection of magnetic field lines creates the conditions for two plasmas of different origins to meet on some boundary. There are some localized effects related to the process of reconnection itself, but even when the reconfiguration of field lines occurs without significant energy release, the process of the interpenetration of two plasmas of different origin lead to important modifications of the particle distributions in some regions of the newly formed flux tube. The interpenetration is of course a time-dependent problem, in which the plasma from the denser side of the boundary will tend to flow toward the other side. In the course of this process, a rarefaction front will propagate into the denser of the two plasmas, while a compression front will propagate into the sparser one. Between these two fronts lies the interpenetration region, where both plasmas are mixed, whereas beyond the fronts the plasmas are still essentially unperturbed concerning their initial state. The purpose of this section is to provide a kinetic description of this rarefaction/compression structure; we shall do so by following the one-dimensional treatment of \citep{Gurevich1968}.
    
\subsection{Interpenetration of two collisionless neutral gases}
\label{section:neutral_gases}
We first discuss the problem of the interpenetration of two collisionless neutral gases with different macroscopic parameters (e.g. density, temperature, or pressure). In general, the gas particles may be described by the velocity distribution function $F(\textbf{v},\textbf{r},t)$. We consider the one-dimensional problem consisting in two gases initially occupying a half-space each, being separated by an impenetrable boundary having a normal vector along the $x$ axis. The Galilean frame of reference in which the problem is formulated is the one in which this boundary is initially at rest. Let us open the boundary at the initial time, allowing to particles of the two gases to freely move. Since the particles interact neither with each other nor with any external field, the evolution of their velocity distribution is simply given by    
\begin{equation}
\frac{\partial F(v_x,x,t)}{\partial t} +v_x \frac{\partial F(v_x,x,t)}{\partial x}=0
\label{kinetic_collisionless_neutral}
\end{equation}
where $F(v_x,x,t)=\int{dv_{z}}\int{dv_{y}}F(\textbf{v})$ is the reduced velocity distribution along the $x$ axis. The characteristics of this partial differential equation are the curves $v_x = v_{x0}$ and $x-v_xt = x_0$, where $v_{x0}$ and $x_0$ are a particle's velocity and position at $t=0$, so that the general solution of equation~\ref{kinetic_collisionless_neutral} is
\begin{equation}
F(v_x,x,t)=F_0(v_x, x-v_xt),
\end{equation}
where $F_0$ is the distribution at initial time, $F_0(v_x, x) \equiv F(v_x, x,0)$. In our case this initial condition can be written as $F_0(v_x, x<0) = F_1(v_x)$ and $F_0(v_x, x>0) = F_2(v_x)$ -- with $F_1$ and $F_2$ the velocity distributions of the gas particles initially on both sides of the boundary -- and the evolution of the distribution after the opening of the boundary is therefore given as a function of space and time by
\begin{equation}
F(v_x,x,t) = F_1(v_x) \left[ 1-\Theta\left(x-v_xt \right) \right] + F_2(v_x)\Theta\left(x-v_xt \right) 
\label{solution_collisionless_neutral}
\end{equation}

Here, $\Theta(u)$ is the Heaviside step function, which is equal to 1 for positive values of $u$ and 0 for negative values. This solution is self-similar, in the sense that it depends on space and time only through the ratio $x/t$. This was to be expected since the problem treated contains no characteristic length or time scales. To get an intuitive understanding of the behavior of the system, let us consider the distribution at a position $x>0$ and time $t$: the above solution (eq.~\ref{solution_collisionless_neutral}) tells us that it will consist of particles initially in the left half space with velocities greater than $x/t$ (since the other particles initially on the left side could not reach the position $x$ at time $t$), and of particles initially in the right half-space, with velocities less than $x/t$. Indeed, all particles initially in the left half and with velocities greater than $x/t$ will have reached positions greater than $x$ at time $t$.

The simplicity of the solution for the mixing of two non-interacting neutral gases makes it possible to derive useful expressions for the evolution of the first moments of the distribution, which are related to the macroscopic fluid variables describing the gas. We detail this in Appendix~\ref{sec:appendix_neutral_gas}, assuming that $F_1(v_x)$ and $F_2(v_x)$ are Maxwellian distribution functions.

\subsection{Interpenetration of two collisionless quasi-neutral plasmas}

We now consider the problem of the interpenetration of two plasmas. The geometry and notations are the same as in the previous section, but the gases are no more neutral but composed of protons, of charge $M$ and charge $e>0$, interacting with an electric potential $\varphi$. The equation describing the evolution of their reduced velocity distribution function is 
\begin{equation}
\frac{\partial F(v_x,x,t)}{\partial t}  + v_x \frac{\partial F(v_x,x,t)}{\partial x} = \frac{e}{M}\frac{\partial \varphi(x,t)}{\partial x}\frac{\partial F(v_x,x,t)}{\partial v_x}.
\label{kinetic_ion}
\end{equation}
In general, the electric potential must be related through the Poisson equation to both the protons and electrons densities -- the latter being solution of an evolution equation like equation~\ref{kinetic_ion}. This non-linear system is difficult to solve. There are two main ways to simplify it: either one neglects the motion of the protons and essentially studies electrostatic structures at the electron scales, like double layers \citep{Block1978}, or one avoids the description of the motion of the electrons by assuming that it is fast enough to ensure quasi-neutrality in all the plasma. This latter approach was initially developed by \citep{Gurevich1968, Gurevich1966} to describe the expansion of a plasma into a vacuum or into another plasma. The very same approach we will use in our study. 
However, solving the very same one-dimensional equations we consider the particle distributions as three-dimensional. For our study, we shall choose them initially as isotropic Maxwellian but one of the goals of our study is to study the temporal-spatial evolution of these distribution functions.    
The assumption of quasi-neutrality is realistic if all the length scales are large compared to the electron Debye length in the plasma. As will be seen in the following, the length scales in the problem are all bound to increase with time due to the self-similar nature of the expansion. Therefore if it is valid at a given time, the quasi-neutral assumption will always be valid for further times. Since the width of the boundary (i.e. the size of the initial density or temperature gradient) is the only spatial scale appearing in the problem, we can reasonably conclude that the quasi-neutral approximation will be valid for all times if this initial width is large compared to the Debye length. As we discussed above the characteristic scales of the reconnection region are supposed to be comparable with the ion Larmor radius that is significantly larger than the Debye length, thus the quasi-neutrality condition is well justified on these scales as on the heights of the low corona/ chromosphere as $\lambda_d / \rho_l \ll 1$]. 
Under quasi-neutral approximation, the electron population is assumed to be at all times in equilibrium in the potential $\varphi$, so that
\begin{equation}
e\varphi(x,t) = kT_e \ln n(x,t)
\label{electric_potential}
\end{equation}
where the electron population is assumed to be isothermal at temperature $T_e$ (this implies in particular that the electron temperature must initially be the same in the plasmas on both sides of the boundary), and $n(x,t)$ is the electron density. Because of plasma neutrality we assume the electron density to be everywhere equal to that of the proton, $n(x,t) = \int F(v_x,x,t) dv_x$. 

Formally, one should have $\ln \left(n(x,t) / n_0 \right)$ in equation~\ref{electric_potential}. This, however, simply leads to a constant shift in the electric potential, whereas only the derivative of the latter plays a role on the dynamics of the particles.

Equations~\ref{kinetic_ion}-\ref{electric_potential} describe the evolution of the ions and of the potential. Injecting one into the other, one can see that the only intrinsic scale of the problem, $\sqrt{kT_e/M}$, has the dimension of a speed (in the following we shall use the isothermal sound speed defined as $c_s = \sqrt{2kT_e/M}$ to normalize all the dimensioned quantities). Therefore the problem is related to no spatial nor temporal scales (as was already the case in the neutral gas case), and the expansion will be self-similar, i.e., all the functions will depend on space and time only through the ratio $\xi = x/t$. Using this property, we can reformulate equation~\ref{kinetic_ion} as
\begin{equation}
(v-\xi)\frac{\partial F(v_x,\xi)}{\partial \xi} = \frac{e\varphi'(\xi)}{M} \frac{\partial F(v_x,\xi)}{\partial v_x}.
\label{kinetic_ion_self_similar}
\end{equation}
The characteristic curves of this partial differential equation are solutions of the equation
\begin{equation}
\frac{dv_x}{d\xi} = -\frac{e}{M} \frac{\varphi'(\xi)}{v_x-\xi}.
\label{characteristics_self_similar}
\end{equation}

Following \citet{Gurevich1968}, we obtained approximate solutions to our problem by numerically integrating equation~\ref{characteristics_self_similar} for the characteristics in a given potential $\varphi_0(\xi)$. Then we used these characteristics to calculate the proton density $n(\xi)$, and, from equation~\ref{electric_potential}, the new potential $\varphi_1(\xi)$. Then we iterated by computing the characteristics in the given potential $\varphi_1(\xi)$, and used them to calculate the proton density and, consequently, the new potential $\varphi_2(\xi)$. We continued iterating until the solution got stable, i.e. the potential $\varphi(\xi)$ did not vary by more than a few percent from one iteration to the next one. Of course, we need an ad-hoc function $\varphi_0(\xi)$ to start the iterative process. We used for this purpose the value of $n(x,t)$ given by the exact solution, derived in Appendix~\ref{sec:appendix_neutral_gas}, of the neutral gas expansion.\\

\subsection{Discussion of the solutions to the interpenetration problem}

The iterative method described in the previous section, together with the analytical solution presented in Appendix~\ref{sec:appendix_neutral_gas}, were used to reach the solution $F(v, \xi)$ to the problem of the kinetic expansion of a plasma (or a neutral gas) into another one. Here we present the properties of such a solution, obtained using initial conditions that are relevant to the study of the reconnection of a closed loop with a flux tube opening into the interplanetary medium. We consider the closed field line to be associated with a magnetic structure such as a magnetic arc. The height at which the reconnection processes can then occur are of the order of the size of the arc. The plasma in that region is expected to have a low beta, although there is a severe lack of direct observations at such small scales. The referee drew our attention to the only direct observation of the large density variations over distances of few thousand km in the low corona at heliocentric distances of $1.2 R_{sun}$. \citet{Raymond2014} found observing the sun-grazing Comet Lovejoy large density variations factor-of-6.

In addition, the lower corona and the chromosphere are highly structured, which means that adjacent open and closed structures might differ in their density by much more than an order of magnitude.

The low value of beta implies that the pressure is imposed by magnetic field variations. These are known to be very structured in the vicinity of CH boundaries, as shown in simulations of the magnetic field configuration based on direct measurements of the photospheric magnetic field \citep{Yeates2018, Linker2017}. It is worth mentioning that the predictions of the MHD-based synthetic models of the corona \citep{Linker2017} are found to be in good agreement with the magnetic topology as observed during eclipse images. This justifies the core idea that the difference in particle density between adjacent closed and open magnetic tubes may be much larger than 10. 

The plasma density is known to be higher in  regions of the closed field lines in the chromosphere and in the lower corona. In our model, we suppose that the electron temperature is constant and identical for both plasmas in the whole reconnected flux tube. In the simulation that will be presented below, the temperature is taken to be $T_e =$\si{1}{MK}. The protons, which are originally supposed to be in the closed loop, have a temperature $T_1 = $\si{2}{MK}, while the protons that are in open field lines have a temperature of $T_2 =$\si{1}{MK}. The ratio of plasma densities between regions of closed versus open lines is chosen to be $n_1/n_2 = 5$. The plasma on closed lines is assumed to be initially at rest with respect to the boundary where reconnection occurs, $u_1=0$. On the contrary, the plasma on open field lines is chosen to have a small drift speed toward this boundary, with $u_2 = $ \si{- 30}{km/s} $\approx -0.23$ $c_s$.

All the figures presented below present the physical quantities in dimensionless units. The normalization factors are, for particles densities, the value of the unperturbed density $n_1$ of the denser plasma (initially on the left of the boundary), for velocities the sound speed defined as $c_s = \sqrt{2kT_e/M}$. The pressure is expressed in units of $n_1kT_e$, the particle flux in units of $n_1 c_s$ and the energy and heat fluxes in units of $n_1 c_s kT_e$.\\

\begin{figure}[ht]
\centerline{\includegraphics[width=11cm]{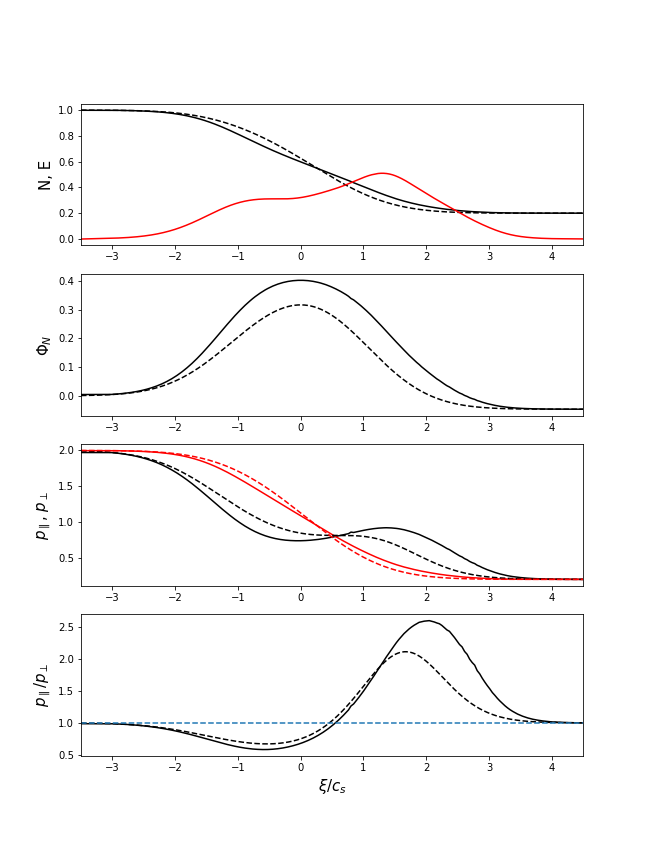}}
\caption{Evolution of the moments of the distribution of the plasma (solid) or neutral gas (dashed) particle velocities as a function of the normalized self similar parameter $\xi/c_s$. From top to bottom are shown the density of the particles $n$ (and $t E \equiv -\varphi'(\xi)$ in red), the particles flux $\Phi_N = n\left< v\right>$, the parallel and perpendicular pressure (black and red, respectively, on the third panel) and the ratio of the parallel to the perpendicular pressure $p_\parallel / p_\perp$. The parameters defining the initial condition are the same as in Figure~\ref{f_parallel_evolution_1}, with the additional $T_{\perp 1} = T_{\parallel 1}$ and $T_{\perp 2} = T_{\parallel 2}$.}
\label{moments_evolution_1}
\end{figure}

Figure~\ref{moments_evolution_1} shows the evolution of the macroscopic moments of the particles' velocity distributions as a function of the normalized parameter $\tau = \xi/c_s$. It illustrates the structure of the interpenetration region. On the left ($\tau \rightarrow -\infty$) is the plasma labeled 1, initially in the closed field lines region, whereas on the right ($\tau \rightarrow \infty$) is the open field lines region.

The top panel shows the density gradient from the dense to the sparse region. It shows the propagation of the rarefaction/compression fronts, which we observe to be located between approximately $-2 < \tau < 2$, which implies that the rarefaction front propagates into the closed field region with a speed of approximately $-2 c_s$, while the compression front propagates into the open field lines with a speed of approximately $2 c_s$. These values are essentially determined, as can be seen on equation~\ref{neutral_density}, by the largest thermal speed $w_{\textrm{max}}$ of the two plasmas, $\tau_{\textrm{front}} \sim \pm w_{\textrm{max}}$. One can observe a weak asymmetry between the position of the two fronts, even in the case where the electric field is neglected (dashed line), which stems from the inclusion of a small drift speed for the plasma with boundary condition $\tau \rightarrow \infty$. 

Finally, by comparing the dashed to the solid curve, we see the effect of the electric field, which accelerates particles from the denser to the sparser plasma, and therefore smoothens the gradient and slightly shifts  the position of the compression front toward higher values of $\tau$. The effect is weak but certainly not negligible, and the neutral solution provides a quite good approximation to the plasma solution for the range of parameters chosen in this simulation. A qualitative analysis shows that the effect of the electric field is strongly dependent upon the ratio of electron to ion temperature and may strongly increase if this ratio becomes large. The red curve of the top panel shows $\varphi'(\xi)$, i.e. the derivative of the electric potential Eq.~(\ref{electric_potential}) with respect to $x/t$. Since the electric field $E(x,t)$ is given by the spatial derivative $\partial \varphi(x,t) / \partial x$, the quantity $\varphi'(\xi) = t E(x/t)$. At a given time $t$, it is thus proportional to the electric field, and shows the profile of the force field in which the ions are evolving.

Figure~\ref{moments_evolution_1} shows the structure of the interpenetration region in terms of higher moments of the distribution function. Notice how the macroscopic particle current directed from the denser to the sparser plasma exists in the interpenetration region. Neglecting the effect of the electric field, and the drift speed compared to the thermal speeds, one can roughly evaluate the maximum value of this flux, which occurs at $\tau \rightarrow 0$ (i.e., for large enough times), to be
\begin{equation}
\Phi_{N, \textrm{max}} =  n_1 \sqrt{ \frac{kT_1}{2\pi M} } - n_2 \sqrt{ \frac{kT_2}{2\pi M}},
\end{equation}
which is just the random particle flux density through the initial boundary. Here again, we can see that the effect of the electric field, accelerating the protons toward the sparser region, is to enhance the particle flux density with respect to the neutral case. 

Finally, the two bottom panels show the evolution of the pressure tensor components $p_\parallel$ and $p_\perp$ and of the pressure anisotropy $p_\parallel/p_\perp$. The plasmas on both sides of the boundary are initially isotropic, with different temperatures. One can see that pressure anisotropies tend to develop in the mixing region, with a perpendicular pressure larger than the parallel one in the wake of the rarefaction front ($\tau < 0$) and the opposite behavior ($p_\parallel/p_\perp>1$) in the compression region $\tau >0$. These anisotropies can be better explained when looking in more detail into the particle distribution functions for different values of $\tau$.

\begin{figure}[h!]
\centerline{\includegraphics[width=13cm]{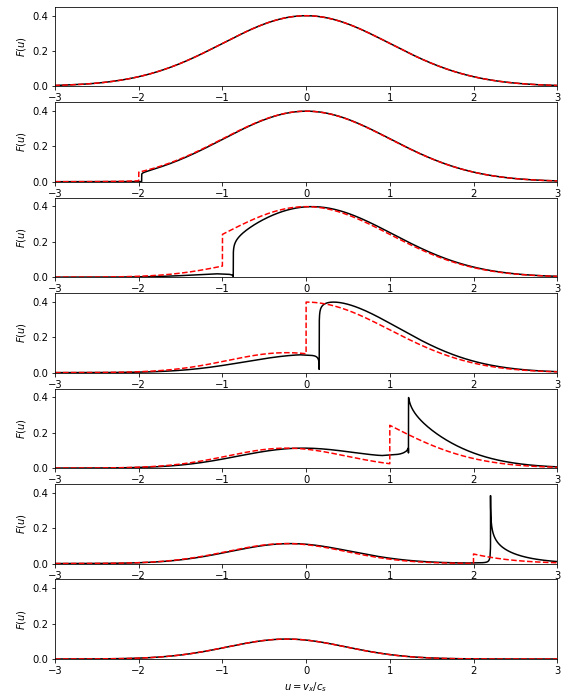}}
\caption{Evolution of the reduced distribution function $F(v_x)$ for the interpenetration of two plasma (solid black) or two neutral gases (dashed red). The distributions are plotted for integer values of the normalized self similar parameter $\tau = \xi / c_s$ ranging from $-3$ (top panel) to $+3$ (bottom panel). Parameters for the initial distributions are $n_2/n_1 = 0.2$, $u_1 = 0$, $u_2 = -0.3 c_s$, $T_{\parallel 1}/T_e = 2$, $T_{\parallel 2}/T_e = 1$.}
\label{f_parallel_evolution_1}
\end{figure}

Figure~\ref{f_parallel_evolution_1} shows the evolution of the reduced velocity distribution functions in space and time, integrated over velocities perpendicular to the direction of the spatial inhomogeneity. The distributions, for the neutral gas and the plasma, are plotted as a function of the normalized velocity $u = v_x / c_s$ for values of $\tau=\xi/c_s = -3, -2, ..., +3$ from top to bottom, so that their shape is shown in different places of the mixing region. On the top panel, one then sees the nearly unperturbed distribution function of the gas initially occupying the left half-space, and on the bottom of the gas initially occupying the right half-space. In between one can observe a very characteristic effect of the "ballistic" mixing on the space-time evolution of the distribution: the formation of a particles beam on the side of the boundary initially occupied by the less dense of the two plasmas, that is, in the wake of the compression front. Comparing the dashed and full curves, one can see that the electric field, by accelerating the protons toward the positive $x$ region, tends to amplify this effect and produce faster beams and denser beams. The presence of these particle beams explains the strong pressure anisotropy observed in the parallel direction on the bottom panel of Figure~\ref{moments_evolution_1} for $\tau > 0$.

\begin{figure}[h!]
\centerline{\includegraphics[width=10cm]{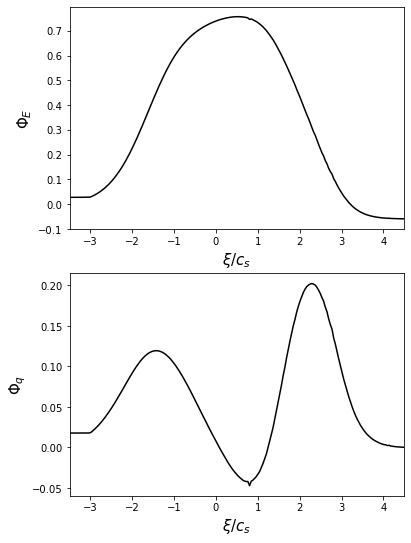}}
\caption{Third order moments of the velocity distribution functions, uncentered (top panel, energy flux density $\Phi_E$) and centered (bottom panel, heat flux density, $\Phi_q$).}
\label{moments_energy_fluxes_1}
\end{figure}

Figure~\ref{moments_energy_fluxes_1} shows the evolution of the energy flux density carried by the protons. It is of course directed from the denser to the sparser regions, and one can see that the electric field tends to enhance it with respect to the neutral gas case. Net energy flux is therefore injected from the closed regions to the open field lines after the reconnection between the flux tubes has occurred. The bottom panel shows the third centered moment, which is the heat flux density carried by the protons. However, this evaluation neglects the heat carried by the electron population, which may be expected to be dominant. It must therefore be an important underestimation of the actual heat flux density.

\begin{figure}[h!]
	\resizebox{\hsize}{!}
	{\begin{tabular}{ccc} 
		  \includegraphics{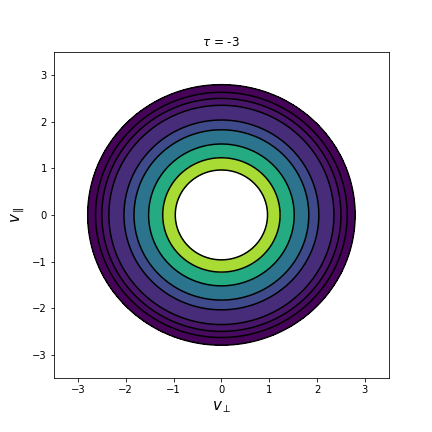} & \includegraphics{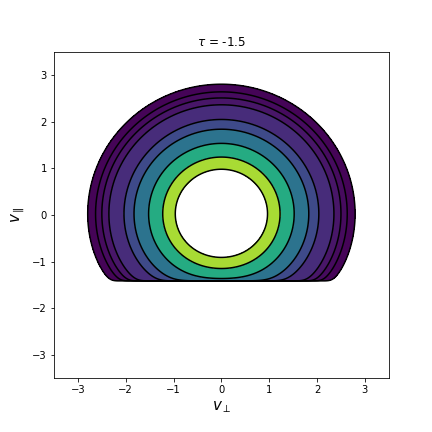} & 
        \includegraphics{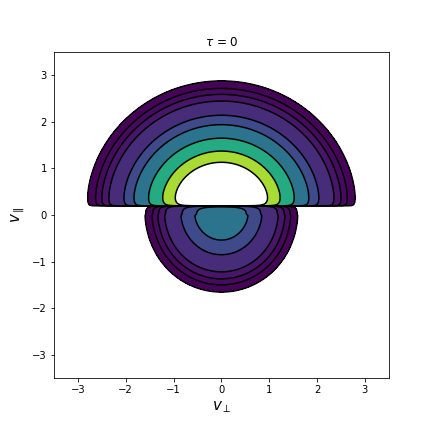} \\
	  \includegraphics{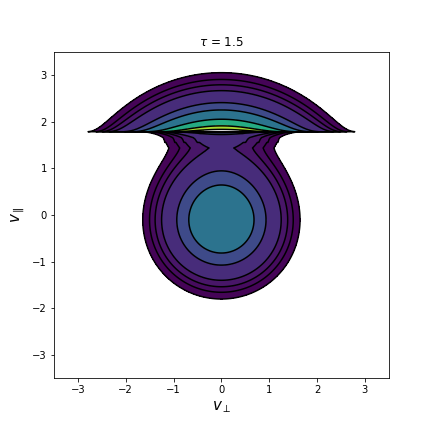} & \includegraphics{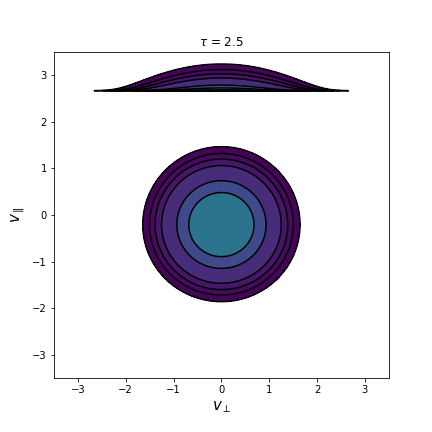} & 
        \includegraphics{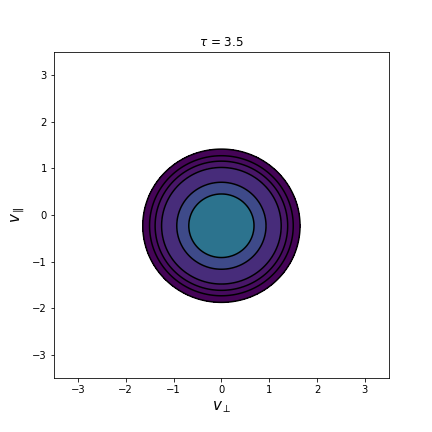} \\
	\end{tabular}
	}
\caption{Evolution of the ion distribution functions along the interpenetration region. }
\label{F_distrub_2d_evolution}
\end{figure}

Figure~\ref{F_distrub_2d_evolution} shows the evolution of the ion distribution function of the protons in the two-dimensional space $(v_\parallel, v_\perp)$. It illustrates the evolution of the cold Maxwellian distribution far on the right of the system ($\tau \gg 1$) to the again Maxwellian, but hot and denser on the left part of it ($\tau \ll 1$). It is well seen that in the intermediate region, there clearly appears double-peak distributions corresponding to the core and beam-like features.

\begin{figure}[h!]
\centerline{\includegraphics[width=10cm]{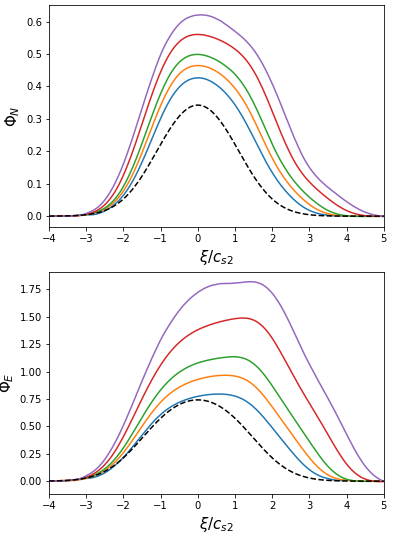}}
\caption{Upper panel: an evaluation of the particle fluxes versus the ratio of the electron to cold ion temperature. $T_e/T_{icold} = $ 1.0 (blue); 1.5 (orange); 2.0 (green); 3.0 (red); 4.0 (violet). Lower panel : energy fluxes, with the same color code. On both panels the dotted lines show the particle and energy fluxes for the interpenetration of two neutral gas.}
\label{Fluxes_vs_Electric_Field}
\end{figure}

In order to illustrate the role of the electric field in energy balance, we carried out several simulations with the different ratios of the electron temperature to ion temperature of the cold population of ions.  In Figure~\ref{Fluxes_vs_Electric_Field} we present an evaluation of the flux of particles (lower panel) and energy (upper panel) for five different temperature ratios $T_e/T_{icold} = 1.0; 1.5; 2.0; 3.0; 4.0$. In order to provide a fair depiction of the effect of the variation of the electron temperature, the fluxes and $\xi$ are not, in this figure, normalized using the electron temperature, but using the proton temperature on the right side of the boundary, that we call $T_{icold} \equiv T_2$, since it corresponds to the open field region, containing the relatively sparse and cold plasma. Thus, $\xi$ is here expressed in units of $c_{s2} = \sqrt{2kT_{icold}/M}$, and the fluxes $\Phi_N$ and $\Phi_E$ in units of $n_1 c_{s2}$ and $n_1 c_{s2} kT_2$, respectively. This figure shows that the fluxes may become significantly higher when the electron temperature increases, that in our case corresponds to the increase of the electric field.    

\section{Injection of a coronal loop plasma into open magnetic field lines}

In this section we apply the results of the previous section to our case of interest, i.e. the mixing, after reconnection of two magnetic flux tubes at a coronal hole boundary: one with a dense and hot plasma originating from closed field regions, and one with a sparser and cooler plasma flowing along open magnetic field lines. First and most importantly, we need to determine the timescale on which the description given in the previous section is valid. If both sides of the flux tube were infinite, then the description should be valid for an infinite time. 

However, there are several elements of our system that may limit the validity of our description. We treat our problem as one-dimensional, but it is not, and so far we have ignored the presence of the magnetic field, which may significantly alter the distribution functions and the whole description. When the magnetic field of the flux tubes undergoes an important change, the one-dimensional treatment becomes incorrect. If these changes are slow enough, the distribution functions are modified due to deviation of the velocity vector of particles with respect to the magnetic field. In the case of slow variations, the effect may be treated by making use of energy conservation and adiabatic invariant conservation. Such changes in their turn may lead to instabilities. The evolution of the particle velocity distribution function forms a  positive slope that leads to the generation of electrostatic instabilities. The most important limitation will depend upon plasma parameters, such as magnetic field configuration variations and inhomogeneity of the system. Thus we shall limit ourselves to qualitative estimates, leaving a more detailed study for future computer simulations. The qualitative upper estimate may be obtained by taking the characteristic scale of the spatial variation of the magnetic field $L_{\textrm{mag}}$ and the characteristic velocity of the propagation of the rarefaction front$\sim  c_s$. One can evaluate the characteristic time for the validity to be less than $T \sim L_{\textrm{mag}}/c_s$.

The scale $L_{\textrm{mag}}$ may be determined by the characteristic scale of the magnetic field variations of the loop field, but if the reconnection occurs around the feet of the funnel, it may be less or even significantly less than the size determined by the field variations in the arc.

For $t \gg T$, the process of the injection of plasma into the heliosphere will occur with the newly formed tube and thus the boundary conditions for the plasma in the source region, so $T$ gives us an order of magnitude of the timescale on which the whole transition process takes place, and it can be used to evaluate various quantities of interest.
 First of all, the size of the region where the proton beam and large pressure anisotropies develop, as can be seen on Figure~\ref{moments_evolution_1}, is around $1<\tau<3$, so the proton beam shall, at the end of the injection process, be found in a spatial region with an extension around $2  L_{\textrm{mag}}$. It is also of interest, of course, to estimate the quantity of matter and energy that will be injected into the open field lines. A rough order of magnitude can be obtained by just multiplying the maximum of the functions $\Phi_N$ or $\Phi_E$ by $T$. The total amount of matter and energy injected toward the interplanetary medium per unit of time will then just be these numbers multiplied by the area of the flux tubes reconnected per unit of time.

\section {Discussion and Conclusions}

Reconnection plays a key important role in the formation of slow and moderate solar winds \citep{Gosling1997, Fisk1999, Fisk2001, Fisk2005, Fisk2006} and numerical studies have described the process of re-configuration of the magnetic field due to its development \citep{Edmondson10, Edmondson12}. The major results of these studies may be resumed as follows.

First, IR generally does not change the global structure of the magnetic field in the photosphere and corona; it remains quasi-potential, but leads to the formation of smaller-scale isolated islands of open field lines. The latter are formed on near coronal hole boundaries but may also be immersed deeply in the regions of closed field lines. Thus, the sources of the solar wind may lie deep in the regions of the closed field configurations. This process is crucial for the formation of slow and moderate winds. An additional strong argument in favor of the idea that part of the solar wind magnetic field lines originated in the regions of the closed field configuration is presented in the study of the spatial evolution of the magnetic field from the photosphere to the heliosphere, which shows that the magnetic field distant flux is larger than that produced by coronal holes only \citep{Linker2017}.  

Second, IR occurs at altitudes corresponding to the height of the arcs in the chromosphere and low corona, i.e. regions where the plasma beta is rather low. This implies that the reconnection process occurs in the presence of a quite strong guiding field. Because the plasma is highly structured, the density and temperature of such a reconnecting system may exhibit strong spatial variations. This suggests that such a reconnection, which according to \citet{Antiochos2007} leads to a slow reformation of the magnetic configuration from a quasi-potential global configuration to another, does not provide a large energy release during its development. The amount of energy dissipation remains very small as we have shown in textbf{Introduction}.

Third, a new source of solar wind plasma emerges: the reconfiguration of the magnetic field structure, which connects the lines immersed in the region of closed field lines with open field lines, now allows plasmas coming in the latter to enter the solar wind. This may significantly change the characteristics of the solar wind because this plasma is already pre-heated, and can potentially have quite different density and particle distributions. 

In this study we have concentrated on the consequences of this reconfiguration process, showing that in connecting flux tubes from different origins it creates favorable physical conditions for the emergence of electic potential differences between plasmas in neighboring magnetic tubes. We have shown that this interpenetration of two plasmas of different origins is determined by the temperature and distribution of the electrons. The electron temperature may significantly exceed that of the ions that come from the low corona. However, other mechanisms may generate such potential differences, such as the electromotive "quasi-potential" between the field lines motion at the level of the photosphere. Most importantly, we find that the connection of low-density and cool plasma with a hotter and denser one can lead to a very important increase of the energy flux ejected in the anti-sunward direction. 
    

Our approach is based on the absence of any characteristic scale in the problem of the transient evolution of the interpenetration of two plasmas that are initially separated by the surface at which the main parameters exhibit a discontinuity. This problem is similar to that of the interpenetration of two neutral gases for which the theory of classical hydrodynamics tells us that the solutions are self-similar and depend only on $\xi = x/t$. The solutions we find in addition reveal a very important physical effect, which is ion acceleration around the boundary separating the two plasmas. This effect can be explained by different populations seeking quasi-neutrality. We show that the difference in fluxes and other properties, such as a heat flux, can become very large in the particular case where the electron temperature is much larger than that of the ions. This problem had already been addressed in the 60's and early 70's by \citet{Gurevich1968, Gurevich1966}. These authors studied the one-dimensional problem and found self-similar solutions numerically. Here we used a similar approach, but in addition we considered 3D ion distributions. This extension allowed us to reveal the formation by this interpenetration of special double-peaked ion distributions with strong variations of the degree of anisotropy of the ion distribution function. 

Solving the spatially 1D problem and choosing the initial conditions to be isotropic, one can show that these variations in the degree of anisotropy leads to the emergence of regions in which the perpendicular pressure becomes significantly higher than the parallel one. The opposite can occur in other regions. Since these processes are supposed to occur in the vicinity of the feet of funnels, where the super-expansion of the magnetic field lines occurs \citep{Cranmer2007}, the further evolution of the ion distribution function (assuming the double adiabatic approximation to remain applicable) could lead to favorable conditions for developing plasma anisotropic instabilities, such as the firehose and mirror instabilities. 

The ion distribution function that results from the interpenetration process consists of a core and a beam-like population, which correspond to protons originating from different parts of the initial distribution, i.e. arriving from different sides of the separatrix boundary. This distribution, which is illustrated in Figure~\ref{F_distrub_2d_evolution}, shows patterns that are strikingly similar to the ``hammerhead'' distributions observed by the SWEAP instrument aboard PSP \citep{Verniero2022}. The latter are typically associated with regions adjacent to the Heliospheric Current Sheet, near the open/closed field line boundary, and some are associated with boundaries of switchback-type structures. This further supports the idea that IR, switchbacks and the interpretation of plasmas of different origins, are intimately related.

Let us finally note that another consequence of these dynamics may be the acceleration of minor ions. The particular aspect will be discussed in a forthcoming paper.

\section{Acknowledgements.} 

The authors are grateful to the referee for very useful comments that helped to improve the article. V.K. is grateful to Hugh Hudson, Marco Velli, Didier Mourenas and Domenique Delcourt for useful discussions. V.K., C.F., and T.D. acknowledge the financial support of CNES in the frame of the Parker Solar Probe grant. J.V. acknowledges support from NASA PSP-GI grant 80NSSC23K0208. Parker Solar Probe was designed, built, and is now operated by the Johns Hopkins Applied Physics Laboratory as part of NASA's Living with a Star (LWS) program (contract NNN06AA01C).

\newpage
\appendix
\section{Appendix: Calculation of the first moments for the collisionless neutral gas expansion}
\label{sec:appendix_neutral_gas}
We calculate in this appendix the first moments of the distribution function of two interpenetrating neutral gas. As seen in the section \ref{section:neutral_gases}, the distribution function takes the general form given by equation~\ref{solution_collisionless_neutral}. In this equation, $F_1(\textbf{v})$ and $F_2(\textbf{v})$ are the velocity distribution functions at time $t=0$ of the gases in the $x<0$ and $x>0$ half spaces respectively. Here we consider the case of Maxwellian distributions for $F_1$ and $F_2$, with possibly different temperatures in the direction $x$ (along the normal to the boundary between the gases) and in the directions $y$ and $z$ (that we refer to as the perpendicular direction):
\begin{equation}
F_i(\textbf{v}) = \frac{n_i}{(\sqrt{\pi})^3w_{\parallel,i} w_{\perp,i}^2} \exp \left( -\frac{(v_x - u_i)^2}{w_{\parallel,i}^2}\right) \exp \left( -\frac{v_y^2 +v_z^2}{w_{\perp,i}^2}\right).
\label{initial_distributions}
\end{equation}
In this expression, $n_i$ is the density of the gas $i$ and the thermal speeds are defined such as $w_{\parallel,i} = \sqrt{2kT_{\parallel,i}/M}$ and $w_{\perp,i} = \sqrt{2kT_{\perp,i}/M}$. $M$ is the mass of the particles and $u_i$ is the mean drift velocity of the gas $i$ with respect to the boundary at time $t=0$.\\

The macroscopic variables associated to the motion of the gas are defined as the statistical moments of the distribution (\ref{solution_collisionless_neutral}). Since the calculation frequently appears truncated moments of the Maxwellian functions defined above, it is convenient to define the following integrals
\begin{equation}
I_1^{(n)}(z) = \int_z^{\infty} \frac{n_1}{\sqrt{\pi}w_{\parallel, 1}}\exp \left( \frac{v_x^2}{w_{\parallel,1}^2}\right) v_x^n dv_x
\end{equation}
and
\begin{equation}
I_2^{(n)}(z) = \int_{-\infty}^{z} \frac{n_2}{\sqrt{\pi}w_{\parallel, 2}}\exp \left( \frac{v_x^2}{w_{\parallel,2}^2}\right) v_x^n dv_x.
\end{equation}
These integrals can be expressed in terms of usual functions. Since we are only interested in calculating the 3 first moments, we only give expressions up to order $n=3$:
\begin{equation}
I_1^{(0)}(z) = \frac{n_1}{2} \left[ 1-\textrm{erf} \left(\frac{z}{w_{\parallel,1}} \right) \right]
\label{I_10}
\end{equation}
\begin{equation}
I_2^{(0)}(z) = \frac{n_2}{2} \left[ 1+ \textrm{erf} \left(\frac{z}{w_{\parallel,2}} \right) \right]
\label{I_20}
\end{equation}

\begin{equation}
I_1^{(1)}(z) = \frac{n_1 w_{\parallel,1}}{2\sqrt{\pi}} \exp \left(-\frac{z^2}{w_{\parallel,1}^2} \right)
\end{equation}
\begin{equation}
I_2^{(1)}(z) = - \frac{n_2 w_{\parallel,2}}{2\sqrt{\pi}} \exp \left(-\frac{z^2}{w_{\parallel,2}^2} \right) 
\end{equation}

\begin{equation}
I_1^{(2)}(z) = \frac{n_1 w_{\parallel,1}^2}{4\sqrt{\pi}} \left[ \sqrt{\pi} \left(  1-\textrm{erf} \left(\frac{z}{w_{\parallel,1}} \right)  \right) + 2 \frac{z}{w_{\parallel,1}}\exp \left( -\frac{z^2}{w_{\parallel, 1}^2}\right) \right]
\end{equation}
\begin{equation}
I_2^{(2)}(z) = \frac{n_2 w_{\parallel,2}^2}{4\sqrt{\pi}} \left[ \sqrt{\pi} \left(  1+\textrm{erf} \left(\frac{z}{w_{\parallel,2}} \right)  \right) - 2 \frac{z}{w_{\parallel,2}}\exp \left( -\frac{z^2}{w_{\parallel, 2}^2}\right) \right]
\end{equation}

\begin{equation}
I_1^{(3)}(z) = \frac{n_1 w_{\parallel,1}}{2\sqrt{\pi}} \left( w_{\parallel,1}^2 + z^2 \right) \exp \left(-\frac{z^2}{w_{\parallel,1}^2} \right) = \left( w_{\parallel,1}^2 + z^2 \right) I_1^{(1)}(z)
\end{equation}
\begin{equation}
I_2^{(3)}(z) = - \frac{n_2 w_{\parallel,2}}{2\sqrt{\pi}} \left( w_{\parallel,1}^2 + z^2 \right) \exp \left(-\frac{z^2}{w_{\parallel,2}^2} \right) =\left( w_{\parallel,2}^2 + z^2 \right) I_2^{(1)}(z)
\end{equation}

Using these definitions it is now easy to find the expressions for the moments of interest for our problem. In order to make the notations a bit more compact, we introduce the notation $I_i^{(n)} \equiv I_i^{(n)} \left(z= x/t - u_i \right)$. We obtain for the density
\begin{equation}
n(x,t) = \int F(x,\textbf{v},t) d\textbf{v} = I_1^{(0)} + I_2^{(0)}.
\label{neutral_density}
\end{equation}
The particle flux density is given by $\Phi_N(x,t) = \int F(x,\textbf{v},t) v_x d\textbf{v}$, so that
\begin{equation}
\Phi_N(x,t) =  I_1^{(1)}+ I_2^{(1)} + u_1 I_1^{(0)} + u_2 I_2^{(0)}.
\label{neutral_particle_flux}
\end{equation}
The stress tensor, defined as $\Pi_{\alpha\beta} = \int F(x,\textbf{v},t) M v_\alpha v_\beta d\textbf{v}$ is obvisously diagonal, since the distribution considered is centered in $v_y$ and $v_z$. Defining the components $\Pi_{xx} = \Pi_\parallel$ and $\Pi_{yy} = \Pi_{zz}  = \Pi_\perp$, one has
\begin{equation}
\Pi_\perp =  kT_{\perp,1} I_1^{(0)} +  kT_{\perp,2} I_2^{(0)}
\end{equation}
and
\begin{equation}
\Pi_\parallel = M\left( I_1^{(2)} +  2 u_1 I_1^{(1)} + u_1^2 I_1^{(0)} + I_2^{(2)} +  2 u_2 I_2^{(1)} + u_2^2 I_2^{(0)} \right)
\end{equation}
The pressure tensor components $p_{\alpha\beta}$ are defined with respect to $\Pi$ such as $p_{\alpha\beta} = \Pi_{\alpha\beta} - nM \left<v_{\alpha} \right> \left<v_{\beta} \right>$. The expanding gas has no mean velocity in the $y$ and $z$ directions, and the perpendicular pressure $p_\perp$ is therefore equal to $\Pi_\perp$. In the parallel direction, the pressure is $p_\parallel = \Pi_\parallel - M\Phi_N^2/n$ where $\Phi_N$ is the particle flux given by equation~\ref{neutral_particle_flux} and $n$ the density given by equation~\ref{neutral_density}. 

Finally the energy flux density,  defined as $\Phi_E(x,t) = \int F(x,\textbf{v},t) \frac{1}{2}M\textbf{v}^2 v_x d\textbf{v}$, is given by
\begin{eqnarray}
\Phi_E(x,t) & = 2 k T_{\perp,1} \left( I_1^{(1)} + u_1 I_1^{(0)} \right)+  2 k T_{\perp,2} \left( I_2^{(1)} + u_2 I_2^{(0)} \right) + \nonumber \\
& \frac{1}{2} M \left( I_1^{(3)} + 3 u_1 I_1^{(2)} + 3 u_1^2 I_1^{(1)} + u_1^3 I_1^{(0)} + \right. \\
& \left. I_2^{(3)} + 3 u_2 I_2^{(2)} + 3 u_2^2 I_2^{(1)} + u_2^3 I_2^{(0)} \right)  \nonumber
\end{eqnarray}


\bibliography{addon}
\bibliographystyle{aasjournal}

\end{document}